\documentclass[aps,prb,twocolumn,floats]{revtex4}

\usepackage{amssymb}
\usepackage{amsmath}
\usepackage{graphicx}
\usepackage{bm}
\usepackage{mdframed}
\usepackage{color}
\usepackage{float}

\begin{document}

\bibliographystyle{prsty}
\author{Ricardo Zarzuela and Yaroslav Tserkovnyak}
\affiliation{Department of Physics and Astronomy, University of California, Los Angeles, California 90095, USA}

\begin{abstract}
We investigate the formation and dynamics of spin textures in antiferromagnetic insulators adjacent to a heavy-metal substrate with strong spin-orbit interactions. Exchange coupling to conduction electrons engenders an effective anisotropy, Dzyaloshinskii-Moriya interactions, and a magnetoelectric effect for the N\'{e}el order, which can conspire to produce nontrivial antiferromagnetic textures. Current-driven spin transfer enabled by the heavy metal, furthermore, triggers ultrafast (THz) oscillations of the N\'{e}el order for dc currents exceeding a critical threshold, opening up the possibility of Terahertz spin-torque self-oscillators. For a commonly invoked antidamping-torque geometry, however, the instability current scales with the energy gap of the antiferromagnetic insulator and, therefore, may be challenging to reach experimentally. We propose an alternative generic geometry for inducing ultrafast autonomous antiferromagnetic dynamics.
\end{abstract}


\title{Antiferromagnetic textures and dynamics on the surface of a heavy metal}

\maketitle 

{\it Introduction}.|Antiferromagnetic spin textures produce minimal stray fields, are robust against electromagnetic perturbations, and display ultrafast spin dynamics, three features that make them attractive as potential active elements in next-generation spin-transport and memory-storage devices.\cite{AFM-spintronics} Recent years have witnessed a growing interest in the inherently antiferromagnetic (spin) transport properties.\cite{AHE,SSE,AFM-superfluidity,AFM-T} However, the N\'{e}el order is relatively hidden from electromagnetic fields and, therefore, generally not easy to drive or read out. In this regard, spin-transfer torques are well suited to trigger antiferromagnetic excitations\cite{STT1,STT2} and may be as effective for this purpose as in ferromagnets.\cite{AFM-T,Cheng-PRL2014} It appears particularly attractive to manipulate the staggered order parameter through the spin Hall effect. In the usual antidamping-torque geometry,\cite{Cheng-PRL2016a} however, the effective spin accumulation induced by the spin Hall effect must overcome the large gap in the antiferromagnetic spectrum (typically in the range of THz for common materials), translating into prohibitively large charge currents. Therefore, further insights concerning the antiferromagnetic equilibrium states and their spin Hall induced dynamics, in the presence of strong spin-orbit interactions, are desired for further progress.
  
In this Rapid Communication, we construct a phenomenological theory for antiferromagnetic insulators subjected to spin exchange and spin-orbit coupling with an adjacent heavy metal. We focus on energy terms that favor spin textures, with an eye on nontrivial topologies. Furthermore, we study the N\'{e}el order driven out of equilibrium by spin-transfer torques and find the thresholds for current-driven magnetic instabilities for several scenarios, classifying the ensuing nonlinear dynamics. Our primary interest here is in ultrafast self-oscillations of a uniform staggered order, which can be sustained by feasible charge currents (as in, e.g., the ferromagnetic counterparts).

\begin{figure}[t]
\begin{center}
\includegraphics[width=0.9\linewidth]{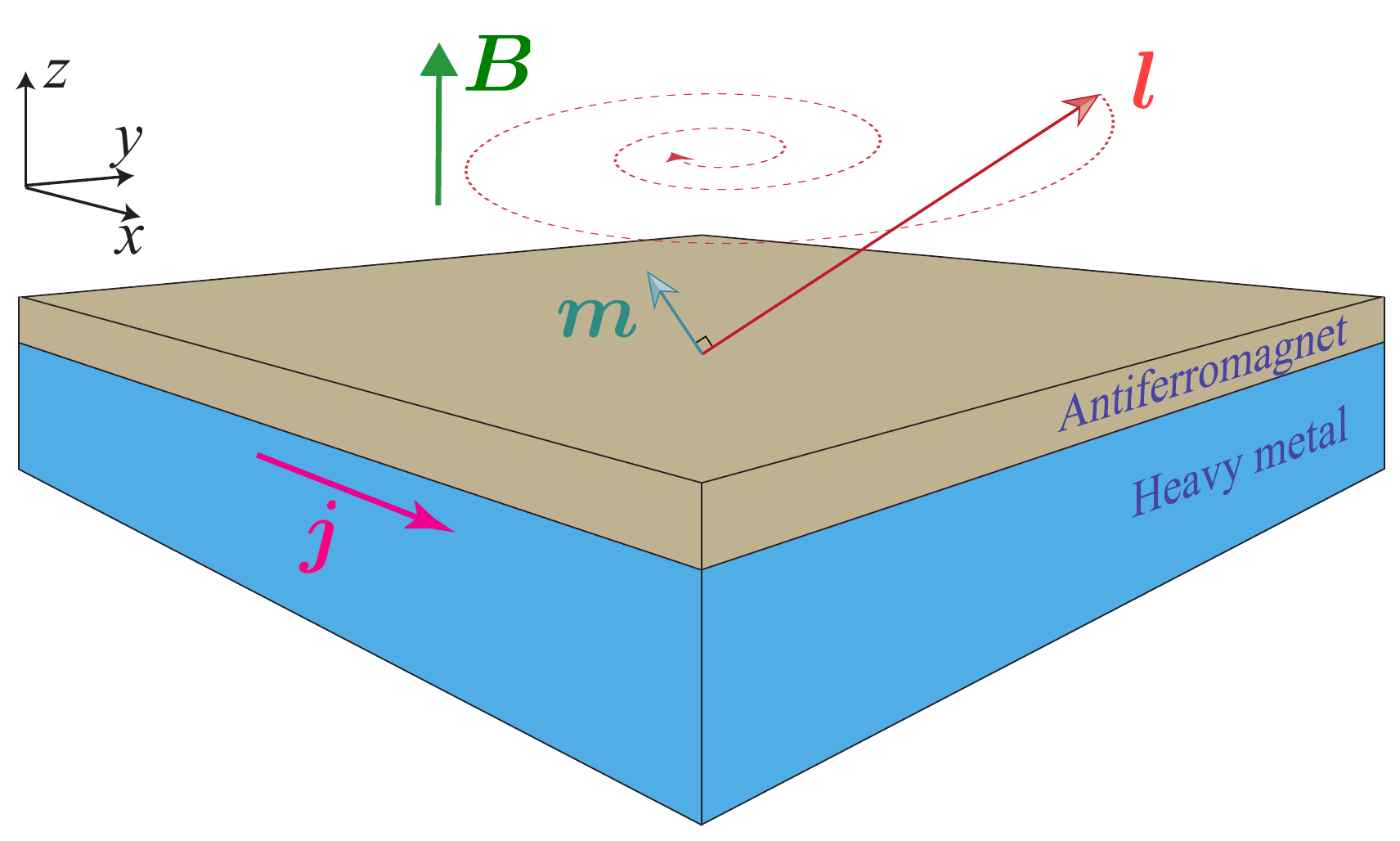}
\caption{A schematic of our heterostructure: An insulating antiferromagnetic film in the N\'{e}el phase is deposited on the surface of a heavy-metal substrate. The (dynamic) magnetic texture is described by the normalized spin density $\bm{m}$ and the N\'{e}el order $\bm{l}$. The coarse-grained field $\bm{j}$ represents the charge-current density flowing within the substrate. The dashed spiral line illustrates dynamics of the N\'{e}el order.}
\label{Fig1}
\end{center}
\end{figure}

{\it Effective theory}.|We regard the heterostructure as a quasi-two-dimensional (2D) system along the {\it xy} plane, which we take to be isotropic at the coarse-grained level, see Fig.~\ref{Fig1}. The reflection symmetry along the $z$ axis is structurally broken and the time-reversal symmetry is also broken due to the existence of the N\'{e}el phase. We focus on bipartite antiferromagnetic insulators, where the two spin sublattices can be transformed into each other by a space-group symmetry of the crystal, and we restrict ourselves to smooth and slowly-varying spin textures. An effective long-wavelength theory for this class of antiferromagnets can be developed in terms of two continuum coarse-grained fields: the (staggered) N\'{e}el field $\bm{l}$ and the normalized spin density $\bm{m}$.\cite{AFM} These fields satisfy the nonlinear local constraints $\bm{l}^{2}\equiv1$, $\bm{l}\cdot\bm{m}\equiv0$, and the presence of a well-developed N\'{e}el order implies (at reasonable fields) $|\bm{m}|\ll1$. The corresponding (2D) Lagrangian density in the continuum limit becomes
\begin{align}
\mathcal{L}_{\textrm{AFM}}[t;\bm{l},\bm{m}]&=\mathcal{L}_{\textrm{kin}}[t;\bm{l},\bm{m}]-\mathcal{F}_{\textrm{AFM}}[\bm{l},\bm{m}],\nonumber\\
\mathcal{F}_{\textrm{AFM}}[\bm{l},\bm{m}]&=\frac{\bm{m}^{2}}{2\chi}-\bm{m}\cdot\mathcal{B}+\mathcal{F}_{\textrm{stag}}[\bm{l}],
\label{eq1}
\end{align}
to quadratic order in both $\bm{l}$ and $\bm{m}$, where $\mathcal{F}_{\textrm{AFM}}[\bm{l},\bm{m}]$ denotes the free-energy density of the antiferromagnet, $\chi$ is the (transverse) spin susceptibility, $\mathcal{B}=\gamma s\bm{B}$ represents the normalized magnetic field, and $s$ is the saturated (2D) spin density.\cite{Comm1} The kinetic (Berry-phase) Lagrangian 
\begin{equation}
\label{eq2}
\mathcal{L}_{\textrm{kin}}[t;\bm{l},\bm{m}]=s\,\bm{m}\cdot(\bm{l}\times\partial_{t}\bm{l})
\end{equation}
establishes the canonical conjugacy between $\bm{l}$ and $s\bm{m}\times\bm{l}$. The functional $\mathcal{F}_{\textrm{stag}}[\bm{l}]$ stands for the exchange and anisotropy contributions to the energy of the antiferromagnet. In the case of isotropic exchange and uniaxial anisotropy, we have $\mathcal{F}_{\textrm{stag}}[\bm{l}]=\frac{A}{2}\sum_{\mu=1,2}(\partial_{x_{\mu}}\bm{l})^{2}-\frac{1}{2}K l_{z}^{2}$, where $A$, $K$ are the stiffness and anisotropy constants, respectively. $K<0$ ($K>0$) describes easy (hard) {\it xy} plane. Both $A$ and $\chi^{-1}$ are proportional to $JS^{2}$, with $J$ being the microscopic exchange energy.

Phenomenologically, the exchange coupling of the N\'{e}el order to conduction electrons of the heavy-metal substrate yields the following contributions to the effective energy of the combined system:
\begin{equation}
\label{eq8}
\mathcal{F}_{\textrm{int}}[\bm{l}]=-\frac{K^{\prime}}{2}l_{z}^{2}-L_{1}\,\bm{l}\cdot\bm{E}+\frac{L_{2}}{2}\left[\bm{l}\cdot\nabla l_{z}-l_{z}\nabla\cdot\bm{l}\,\right],
\end{equation}
where $K^{\prime}$, $L_{1}$, and $L_{2}$ are material-dependent phenomenological coefficients and $\bm{E}$ is the static (in-plane) electric field acting on electrons (here, in equilibrium). Notice that, according to the time-reversal symmetry, $L_1$ $(L_2)$ must be an odd (even) function of the out-of-plane component $l_{z}$ of the N\'{e}el order. The first two terms in Eq.~\eqref{eq8} account for an effective axial anisotropy and a magnetoelectric effect for the N\'{e}el order, respectively. The last term can arise due to structural reflection-symmetry breaking at the interface,\cite{Fert-PRL1980} and describes an inhomogeneous Dzyaloshinskii-Moriya interaction.\cite{DMI,Dzyaloshinskii-JETP1964}

In the absence of the electromagnetic fields, $\bm{E}, \bm{B}\rightarrow0$, the above free energy for the N\'{e}el order reproduces that of a ferromagnetic film with the broken reflection symmetry with respect to the basal plane.\cite{Bogdanov-JMMM1994} In particular, a spiral ground state would arise for values of the parameter $L_{2}/\sqrt{AK_{\textrm{eff}}}$ exceeding the critical value $4/\pi$, where $K_{\textrm{eff}}\equiv K+K^{\prime}$ is the effective anisotropy constant.\cite{yt1} As a specific illustrative example, in the Supplemental Material\cite{SUP} we complement our effective theory with microscopic results for the case of a strong three-dimensional topological insulator (TI) as a heavy (semi)metal.  

{\it Nonequilibrium dynamics}.|Undamped Landau-Lifshitz dynamics of the insulating antiferromagnet are described by the Euler-Lagrange equations for the total Lagrangian $\mathcal{L}_{\textrm{AFM}}-\mathcal{F}_{\textrm{int}}$, subject to the local constraints $\bm{l}\,^{2}\equiv1$ and $\bm{l}\cdot\bm{m}\equiv0$.\cite{LoM} A phenomenological approach well suited to incorporate dissipation into these equations considers a Gilbert-Rayleigh function,\cite{Gilbert-1955} whose dominant term is given by $\frac{1}{2}s\alpha_{ij}\dot{l}_{i}\dot{l}_{j}$ in the low-frequency (compared to the microscopic exchange $J$) regime,\cite{kimPRB15} where $\hat{\alpha}$ denotes the Gilbert-damping tensor.\cite{Comm4} The resulting Landau-Lifshitz-Gilbert-type equations read
\begin{align}
\label{eq15}
s\dot{\bm{l}}&=\chi^{-1}\bm{m}\times\bm{l}+\bm{l}\times\mathcal{B}+\bm{\tau}_{l},\\
\label{eq16}
s(\dot{\bm{m}}+\bm{l}\times\hat{\alpha}\dot{\bm{l}}\hspace{0.03cm})&=\delta_{\bm{l}}\mathcal{F}_{\textrm{eff}}\times\bm{l}+\bm{m}\times\mathcal{B}+\bm{\tau}_{m},
\end{align} 
where $\mathcal{F}_{\textrm{eff}}\equiv\mathcal{F}_{\textrm{stag}}+\mathcal{F}_{\textrm{int}}$ is the effective energy and the spin-transfer torques $\bm{\tau}_{l},\bm{\tau}_{m}$ account for the additional, nonequilibrium electric current-induced spin transport across the interface. 

As usual, we can obtain a dynamical equation for the N\'{e}el order alone by solving for $\bm{m}$ according to Eq.~\eqref{eq15} and substituting it in Eq.~\eqref{eq16}. Notice that the effect of the torque $\bm{\tau}_{l}$ is in general reduced relative to $\bm{\tau}_{m}$ by the small parameters $\hbar\omega/J$ (with $\omega$ denoting the characteristic frequency of the antiferromagnetic excitations), which, again,\cite{yt1} is rooted in the smallness of the susceptibility $\chi\propto J^{-1}$. In the spirit of our low-frequency long-wavelength treatment, we, therefore, disregard this spin-transfer torque (i.e., $\bm{\tau}_{l}$) in what follows. 

The spin torque $\bm{\tau}_{m}$ has two (dissipative) components: the first, so-called spin-orbit torque, is rooted phenomenologically in the spin-Hall effect.\cite{SHE} The second is the texture-induced spin-transfer torque,\cite{Hals-PRL2011} which originates in the spin mistracking of conduction electrons (of the heavy metal) propagating in proximity to the N\'{e}el texture.\cite{STextT} According to the structural symmetries of the heterostructure, they have the form:\cite{Tserkovnyak-PRB2014,Hals-PRL2011,Comm5}
\begin{equation}
\label{eq17}
\bm{\tau}_{m}=\vartheta_{2}\bm{l}\times(\hat{e}_{z}\times\bm{j})\times\bm{l}+\vartheta_{3}\bm{l}\times(\bm{j}\cdot\nabla)\bm{l},
\end{equation}
where the coupling constants $\vartheta_{2}$ and $\vartheta_{3}$ depend on the interplay of spin-orbit and spin-transfer physics at the interface.

\begin{figure*}[t]
\begin{center}
\includegraphics[width=0.9\linewidth]{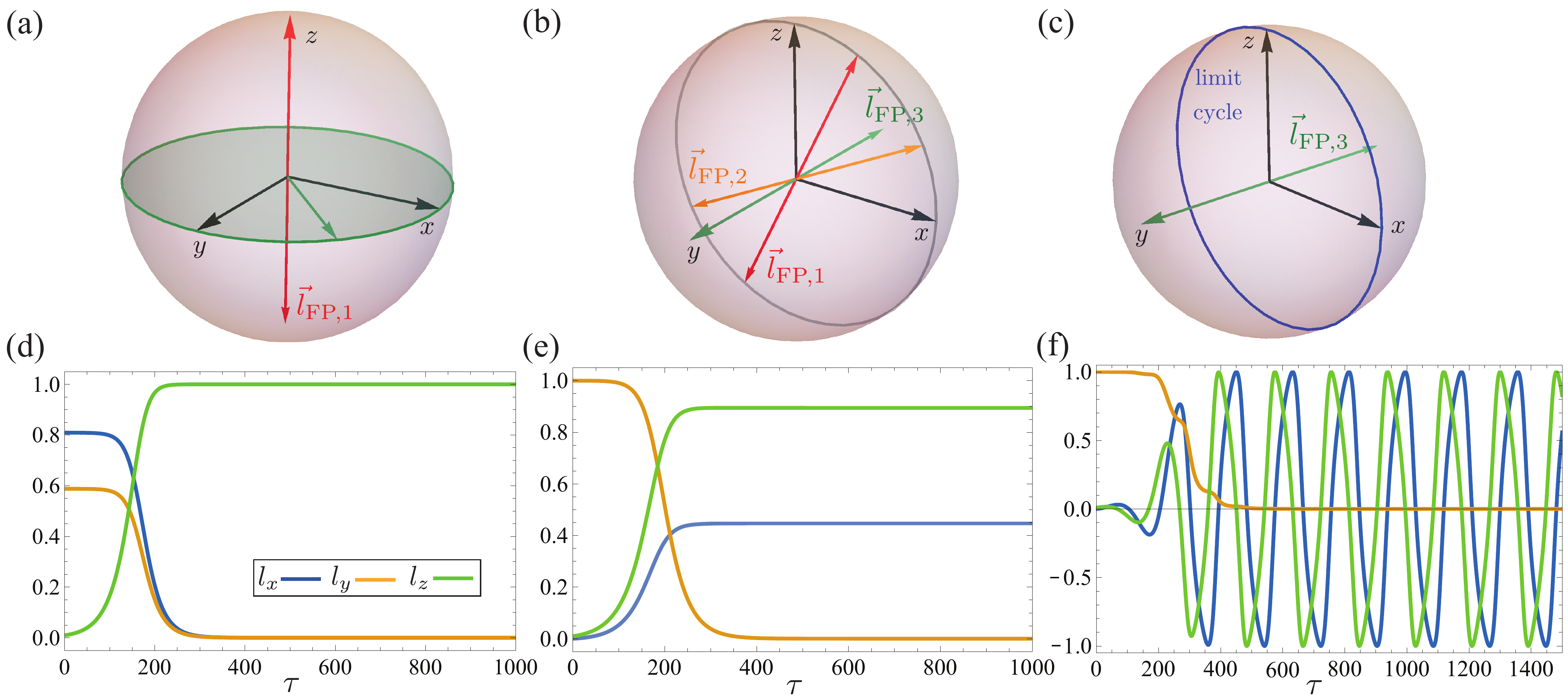} 
\caption{Classification of the current-driven fixed points and limit cycles, according to Eq.~\eqref{eq24}, for (a) $\lambda=0$, (b) $0<\lambda\leq\frac{1}{2}$ and (c) $\lambda>\frac{1}{2}$. Arrows represent fixed points and the blue circle in (c) illustrates the limit cycle in the $xz$ plane. (d-f) Evolution of the components of the N\'{e}el order as a function of time $\tau=t/s\sqrt{\chi}$, setting $K_{\textrm{eff}}=0.01$, $\chi=0.001$ and $\alpha_{\textrm{eff}}=\vartheta_{2}=0.01$. The corresponding critical current is $j_{c}=0.5$. Dynamics are: (d) triggered from the initial point $(0.81,0.59, 0.01)$ at zero current (the limiting fixed point is controlled by the sign of the $z$ component), (e) driven by the current $j=0.4$ from $\bm{l}=(0,0.99,0.01)$, and (f) driven by the current $j=1.2$ from the same fixed point. The periodic time evolution of the components of the N\'{e}el order in the last case leads to the limit cycle plotted in (c).}
\label{Fig2}
\end{center} 
\end{figure*}

Another contribution to the net transfer of spin angular momentum onto the antiferromagnetic texture is provided by the spin-pumping mechanism,\cite{Tserkovnyak-RMP2005} which can be absorbed into the (effective) damping tensor as an interface term, $\hat{\alpha}_{\textrm{eff}}\equiv\hat{\alpha}+\vartheta_{1}/s$.\cite{AFM-T} Here, $\vartheta_{1}$ is the (dissipative) spin-pumping parameter (taken, for simplicity, to be isotropic) related to the spin-mixing conductance of the interface. The combination of Eqs.~\eqref{eq15}-\eqref{eq17} yields the following second-order differential equation for the N\'{e}el order:\cite{Comm6}
\begin{align}
\bm{l}&\times\Big[s^{2}\chi\ddot{\bm{l}}+s\hat{\alpha}_{\textrm{eff}}\dot{\bm{l}}+\delta_{\bm{l}}\mathcal{F}_{\textrm{eff}}+\chi(\bm{l}\cdot\mathcal{B})\mathcal{B}-s\chi\bm{l}\times\dot{\mathcal{B}}\nonumber\\
&-\vartheta_{2}(\hat{e}_{z}\times\bm{j})\times\bm{l}-\vartheta_{3}(\bm{j}\cdot\nabla)\bm{l}\,\Big]-2s\chi(\bm{l}\cdot\mathcal{B})\dot{\bm{l}}=\bm{0},
\label{eq23}
\end{align}
which is the central equation and one of the main results of this Rapid Communication. It is worth mentioning that in order to integrate this equation, it needs to be complemented with the trivial vector identity: $\bm{l}\cdot\ddot{\bm{l}}=(d/dt)^2\bm{l}^2/2-\dot{\bm{l}}^2=-\dot{\bm{l}}^2$, since $\bm{l}^2\equiv1$.

{\it Current-driven monodomain dynamics}.|Magnetic fields may not optimally be suited to manipulate antiferromagnetic textures, as the staggered order suppresses the coarse-grained magnetization in the N\'{e}el phase. In this regard, spin-transfer torques offer an attractive alternative to trigger and control fast antiferromagnetic dynamics, of particular interest being current-induced magnetic instabilities and switching. Starting with the simplest out-of-equilibrium scenario, we consider a uniform state and, therefore, disregard the magnetic torques resulting from the magnetoelectric and inhomogeneous Dzyaloshinskii-Moriya terms in Eq.~\eqref{eq8}, and the texture-induced spin-transfer torque in Eq. \eqref{eq17}. Furthermore, we suppose an easy-$z$-axis anisotropy (i.e., $K_{\rm eff}>0$), absence of an applied magnetic field, and a uniform dc charge-current density injected (without loss of generality) along the $x$ direction, $\bm{j}=j\hat{e}_{x}$. Consequently, Eq.~\eqref{eq23} becomes
\begin{align}
\label{eq24}
s^{2}\chi\ddot{\bm{l}}+s\alpha_{\textrm{eff}}\dot{\bm{l}}&+(\vartheta_{2} j l_{x}-K_{\textrm{eff}}l_{z})\hat{e}_{z}-\vartheta_{2} j l_{z}\hat{e}_{x}\nonumber\\
&+\left[s^{2}\chi\dot{\bm{l}}^{2}+K_{\textrm{eff}}l_{z}^{2}\right]\bm{l}=0,
\end{align}
where, for simplicity, we have taken the full damping tensor to be isotropic.

Stability of the above dynamical system can be analyzed in terms of the parameter $\lambda=\vartheta_{2}j/K_{\textrm{eff}}$: in equilibrium ($\lambda=0$), the fixed points (FPs) are $\bm{l}_{\textrm{FP},1}=\pm\hat{e}_{z}$ along with any $xy$-plane orientation of the N\'{e}el order. See Fig.~\ref{Fig2}(a). From a simple stability analysis,\cite{Hirch} we conclude that $\bm{l}_{\textrm{FP},1}$ are the only stable FPs, and, therefore, any slight out-of-plane perturbation would turn the staggered order from any initial $xy$-plane configuration to a normal direction. See Fig.~\ref{Fig2}(d). When the current is ramped up within the range $0<\lambda\leq\frac{1}{2}$, the set of FPs becomes discrete and reads
\begin{align}
\label{eq25}
\bm{l}_{\textrm{FP},i}&=\pm\left(\frac{\sqrt{2}\lambda}{\sqrt{1-p_{i}\sqrt{1-4\lambda^{2}}}},0,\frac{\sqrt{1-p_{i}\sqrt{1-4\lambda^{2}}}}{\sqrt{2}}\right),\nonumber\\
\bm{l}_{\textrm{FP},3}&=\pm\hat{e}_{y},
\end{align}
where $p_{i}=(-1)^{i}$ and $i=1,2$. Stability theory applied to this case indicates that the $\bm{l}_{\textrm{FP},1}$ are stable FPs whereas the $\bm{l}_{\textrm{FP},2(3)}$ are unstable. See Fig.~\ref{Fig2}(b). Therefore, any slight perturbation acting on $\bm{l}_{\textrm{FP},3}$ will drive the staggered field into one of the fixed points $\bm{l}_{\textrm{FP},1}$. See Fig.~\ref{Fig2}(e). The limiting orientation of the N\'{eel} order depends on the signs of the $x$ and $z$ components of the perturbation. The FPs of Eq.~\eqref{eq24} for $\lambda>\frac{1}{2}$ are $\bm{l}_{\textrm{FP},3}$ and unstable. This leads to the formation of an attractive limit cycle in the $xz$ plane. See Figs.~\ref{Fig2}(c),(f). We thus conclude that the instability threshold of our dynamical system towards self-oscillations is determined by the critical current $j_{c}=\frac{1}{2\vartheta_{2}}K_{\textrm{eff}}$. The frequency corresponding to this limit cycle is in the range of $\omega=\frac{1}{2s}\sqrt{K_{\textrm{eff}}/\chi}$, which agrees with the values obtained from the numerical solution of the full Eq.~\eqref{eq24}.\cite{SUP} It is also instructive to consider a different geometry, in which an in-plane easy-axis anisotropy $K$ is oriented along the $y$ axis (i.e., perpendicular to the direction of the injected current). This is a typical antidamping-torque geometry.\cite{Cheng-PRL2016a,Tserkovnyak-PRB2014} Neglecting $K'$, a precessional instability arises at the critical current $j_{c}^{\star}=\frac{\alpha_{\textrm{eff}}}{\vartheta_{2}}\sqrt{K/\chi}$, and the corresponding antiferromagnetic dynamics have a characteristic frequency of $\omega^{\star}\simeq\frac{1}{s}\sqrt{K/2\chi}$.

The instability thresholds $j_{c}$ and $j_{c}^\star$ scale qualitatively differently with the system parameters, but both appear substantially higher than the typical ferromagnetic instability threshold of\cite{Tserkovnyak-PRB2014} $\sim\frac{\alpha_{\rm eff}}{\vartheta_2}K$ if $\alpha_{\rm eff}\ll1$. Regarding $j_c^\star$, we need to recognize that the quantity $\sqrt{K/\chi}\sim\sqrt{KJ}$ setting the antiferromagnetic resonance frequency is typically much larger than the ferromagnetic resonance frequency, which is governed by $K$ and unaffected by $J$, as the exchange (which is nonrelativistic) is generally stronger than the anisotropy (which is relativistic). Comparing $j_c$ and $j_c^\star$, we thus see that the former scales with the anisotropy, but is not reduced by the damping, as in the ferromagnetic case, while the latter scales with the exchange-enhanced resonance frequency. Note that the scaling of $j_c^\star$ with the energy gap of the antiferromagnet is in agreement with the entropic argument given in Ref.~\onlinecite{Bender-PRL2012} (according to which, the effective spin accumulation induced by the spin Hall effect must overcome the magnon gap). The ratio $j_{c}/j_{c}^\star=\sqrt{K\chi}/2\alpha_{\textrm{eff}}$ is governed by two small parameters: $\sqrt{K/J}$ and $\alpha_{\textrm{eff}}$. In the desirable limit of strong spin-orbit effects, and thus large $\alpha_{\textrm{eff}}$ and $\vartheta_2$, as is the case, for example, in a magnetically-doped TI,\cite{TI-SOC} we may have $j_{c}<j_{c}^\star$.

The phenomenological parameters of our effective theory can be evaluated in a simple diffusive model with weak spin-orbit interactions\cite{Tserkovnyak-PRB2002,Tserkovnyak-PRB2014,Cheng-PRL2016b} as $\vartheta_{1}=\frac{\hbar^{2}}{2a_{M}}\frac{\sigma g^{\uparrow\downarrow}}{h\sigma+2\lambda e^{2}g^{\uparrow\downarrow}\coth(a_{N}/\lambda)}$ and $\vartheta_{2}=\theta_{s}\frac{\hbar e}{a_{M}}\frac{\lambda g^{\uparrow\downarrow}\tanh(a_{N}/2\lambda)}{h\sigma+2\lambda e^{2}g^{\uparrow\downarrow}\coth(a_{N}/\lambda)}$, where $a_{M}$ is the thickness of the antiferromagnetic layer, $g^{\uparrow\downarrow}$ is the spin-mixing conductance (per unit area) of the interface, and $a_{N}$, $\sigma$, $\lambda$, $\theta_{s}$ denote the thickness, conductivity, spin-diffusion length, and the bulk spin Hall angle of the heavy metal, respectively. These expressions coincide with the ferromagnetic case, subject to a generalized understanding of the spin-mixing conductance.\cite{AFM-T,Cheng-PRL2014} An appropriate engineering of the heterostructure (with strong spin-orbit coupling and thin magnetic layer), together with optimizing the switching geometry, are necessary to produce feasible values of the critical currents.

{\it Discussion and outlook}.|Dzyaloshinskii-Moriya interactions in this paper are endowed in the antiferromagnetic insulator by the interface.\cite{Comm7} We have already discussed how our effective theory incorporates an inhomogeneous Dzyaloshinskii-Moriya coupling in response to the proximity of a heavy-metal substrate, giving rise to magnetic superstructures.\cite{Dzyaloshinskii-JETP1964,Bogdanov-JMMM1994} Another possible manifestation of spin-orbit coupling in antiferromagnets is a weak ferromagnetism.\cite{DMI} Whether it is compatible with the sublattice symmetry, $\{\bm{l}\to-\bm{l},\bm{m}\to\bm{m}\}$, depends on the crystallographic structure of the antiferromagnet and its surface. In the Supplemental material\cite{SUP} we illustrate two examples of quasi-2D crystal lattices for which a homogeneous Dzyaloshinskii-Moriya term $\mathcal{F}_{\textrm{DM}}[\bm{l},\bm{m}]=\bm{d}\cdot(\bm{l}\times\bm{m})$ is allowed, where $\bm{d}=d\hat{e}_{z}$ is the Dzyaloshinskii vector along the normal to the interface. Addition of this term to the effective energy $\mathcal{F}_{\textrm{eff}}$ results in a redefinition of the normalized magnetic field $\mathcal{B}\rightarrow\mathcal{B}+\bm{l}\times\bm{d}$ in the equation \eqref{eq1} for the free-energy density. It can be shown, however, that its effect on the antiferromagnetic dynamics at the level of Eq.~\eqref{eq24} can be absorbed by a small shift in the anisotropy constant: $K^{\prime}\rightarrow K^{\prime}-\chi d^{2}$.

Self-oscillations, in the form of limit cycles, are sustained above the critical current $j_{c}$ in the case of the easy-$z$-axis anisotropy.  For the easy-$y$-axis geometry, the nature of the autonomous dynamics beyond the threshold $j_{c}^{\star}$ was shown\cite{Cheng-PRL2016a} to be sensitive to the details of the Gilbert-damping tensor, which can acquire, in particular, an anisotropic form $\propto\bm{l}\times(l_{z}^{2}\dot{\bm{l}}+\dot{l}_{z}\hat{e}_{z})$. The resultant self-oscillation frequencies belong to the THz range for typical insulating antiferromagnets. In order to realize spin-transfer THz auto-oscillators, however, appropriate materials and geometries need to be identified to yield feasible bias currents.

\acknowledgments

This work has been supported by NSF-funded MRSEC under Grant No. DMR-1420451. R.Z. thanks Fundaci\'{o}n Ram\'{o}n Areces for support through a postdoctoral fellowship within the XXVII Convocatoria de Becas para Ampliaci\'{o}n de Estudios en el Extranjero en Ciencias de la Vida y de la Materia.

\newpage

\onecolumngrid
 
\section*{\Large Supplemental Material}
\hspace{0.5cm}

\section{A three-dimensional topological insulator as a heavy (semi)metal}
Strong three-dimensional topological insulators (TIs) represent an extreme case of strong spin-orbit interactions at the interface,\cite{TI-SOC} which relatively easily yields microscopic (model) expressions for the phenomenological parameters of the effective theory. It is worth mentioning that recent advances in electrical gating of magnetically-doped TIs\cite{Fan-NatNano2016} promise great tunability of these coupling coefficients (that govern the magnetic configuration).

We regard the conducting TI surface as a uniform 2D gas of electronic excitations with linearly-dispersing bands,\cite{Comm3} which couple to the antiferromagnetic texture through a local axially-symmetric (single-particle) exchange. This coupling can introduce the following time-reversal symmetry-breaking term into the effective Hamiltonian:\cite{SFN1}
\begin{equation}
\label{eq6}
\hat{\mathcal{H}}_{\textrm{int}}=J_{\parallel}(l_{x}\hat{\sigma}_{x}+l_{y}\hat{\sigma}_{y})+J_{\perp}l_{z}\hat{\sigma}_{z},
\end{equation}
where $J_{\parallel},J_{\perp}$ are the corresponding exchange constants. This term can be absorbed into the Dirac-like description of the surface states,\cite{ExHam} whose integration out\cite{Tserkovnyak-PRBRC2015} yields the following expressions for the phenomenological coefficients:
\begin{equation}
\label{eq9}
L_{1}=J_{\parallel}e\frac{\zeta_{1}+\chi_{1}}{4\pi\hbar v},\quad L_{2}=J_{\parallel}J_{\perp}\frac{\zeta_{2}+\chi_{2}}{4\pi\hbar v},
\end{equation}
where $\zeta_{1},\zeta_{2}$ are offsets reflecting valence-band physics far away from the Dirac point, and
\begin{equation}
\label{eq10}
\chi_{1}(l_{z},\beta,\mu)=\frac{\sinh(\beta J_{\perp}l_{z})}{\cosh(\beta J_{\perp}l_{z})+\cosh(\beta\mu)},\hspace{1cm}\chi_{2}(l_{z},\beta,\mu)=\frac{\sinh(\beta\mu)}{\cosh(\beta J_{\perp}l_{z})+\cosh(\beta\mu)}.
\end{equation}
Here $\beta$, $\mu$ are the thermal factor $1/k_{\textrm{B}}T$ and the chemical potential of Dirac electrons, respectively, and $v$ is the electron speed. According to the time-reversal symmetry considerations, the function $\zeta_{1}$ must be odd in $l_{z}$, whereas $\zeta_{2}$ is an even function (taken constant for simplicity). 

\section{Numerical approach}

Solutions of Eq. (8) are calculated numerically by the following method: we first recast this second-order differential equation as a first-order dynamical system, $\dot{\bm{X}}=F[\bm{X}]$, with $\bm{X}^{\top}=(\bm{l},\dot{\bm{l}}\,)$. Secondly, the trajectories of the N\'{e}el order in the phase space are obtained by integrating this dynamical system with the appropriate initial conditions and subject to the nonlinear constraint $\bm{l}^2\equiv1$. These numerical trajectories are consistent with the fixed points, the limit cycles and the (attractive/repulsive) nearby dynamics predicted by the stability theory. The frequency $\omega$ characterizes the time evolution of the order parameter in a close vicinity of the two points $\bm{l}_{\textrm{FP},3}$, since its analytical expression is obtained from the linearization of the dynamical system at these (unstable) fixed points. It gives, however, a good approximation of the frequency of the self-oscillations of the N\'{e}el order beyond the current threshold $j_{c}$, which is extracted from the long-term dynamics of the numerical solutions. 

\newpage

\section{Homogeneous Dzyaloshinskii-Moriya interaction in films and bilayers}

\begin{figure}[h]
\begin{center}
\includegraphics[width=0.7\linewidth]{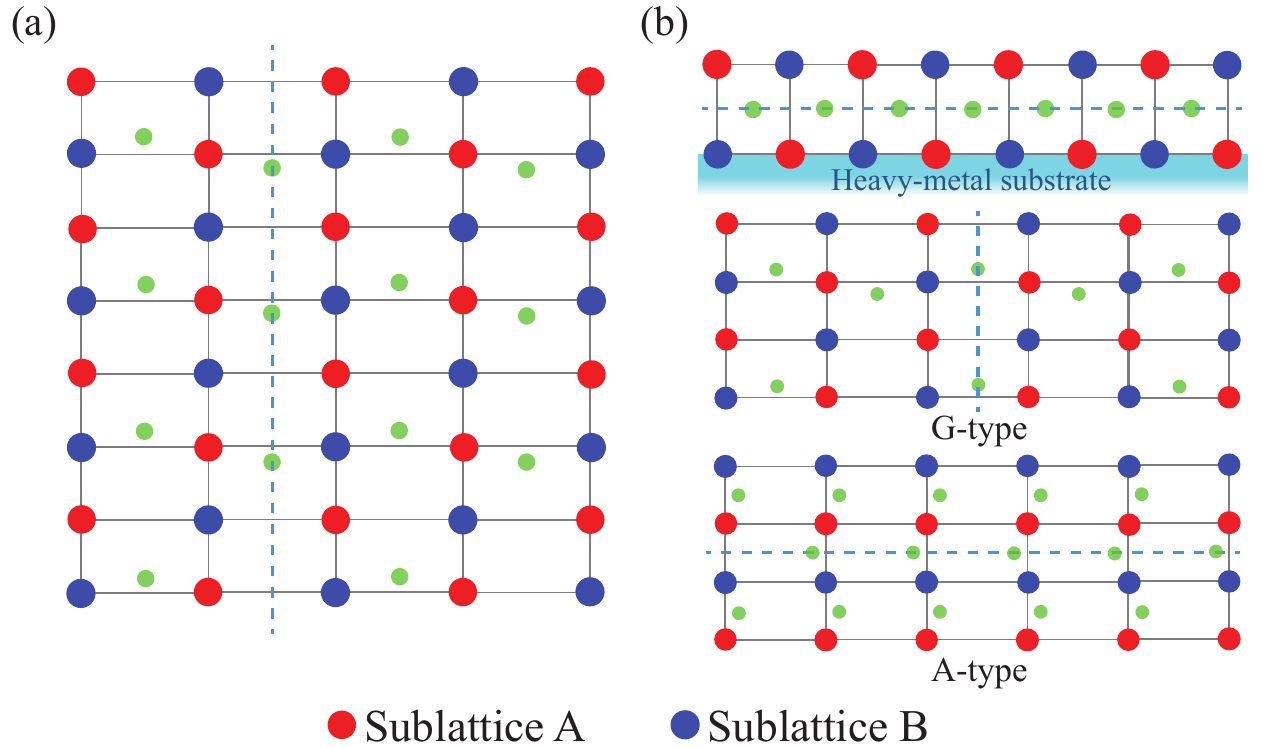}
\caption{Two quasi-2D crystallographic structures compatible with a homogeneous Dzyaloshinksii-Moriya term $\mathcal{F}_{\textrm{DM}}$. (a) Top view of a planar rectangular (or square) lattice and (b) side view (up) and top view corresponding to the G-type (middle) and A-type (down) spin arrangements on a triple-layer rectangular (or square) lattice. Red (blue) circles are magnetic sites belonging to sublattice A (B) of the antiferromagnetic insulator and green circles show (heavy) ions mediating superexchange between the spin sublattices. Dashed lines show reflection-symmetry planes realizing the spin-sublattice symmetry. In (b), the homogeneous Dzyaloshinskii-Moriya interaction is forbidden in the absence of a heavy-metal substrate.}
\label{Fig3}
\end{center} 
\end{figure}

In Fig. \ref{Fig3}, we show two examples of quasi-2D crystal lattices for which a homogeneous Dzyaloshinskii-Moriya term $\mathcal{F}_{\textrm{DM}}[\bm{l},\bm{m}]=\bm{d}\cdot(\bm{l}\times\bm{m})$ is allowed, where $\bm{d}=d\hat{e}_{z}$ is the Dzyaloshinskii vector along the normal to the interface. In the first example, (a), the magnitude $d$ may be amplified by the presence of the heavy-metal substrate. In the second example, (b), no such interaction would exist without the inversion-symmetry-breaking substrate.

\twocolumngrid


\begin{thebibliography}{99}
\bibitem{AFM-spintronics} J. Basset, A. Sharma, Z. Wei, J. Bass and M. Tsoi, Proc. SPIE {\bf 7036}, 703605 (2008); A. H. MacDonald and M. Tsoi, Philos. Trans. R. Soc., A {\bf 369}, 3098 (2011); X. He {\it et al.}, Nat. Mater. {\bf 9}, 579 (2010); D. Sando, A. Barthélémy and M. Bibes, J. Phys. Condens. Matter {\bf 26}, 473201 (2014).
\bibitem{AHE} R. Shindou and N. Nagaosa, Phys. Rev. Lett. {\bf 87}, 116801 (2001); H. Chen, Q. Niu and A.H. MacDonald, {\it ibid.} {\bf 112}, 017205 (2014).
\bibitem{SSE} S. Seki {\it et al.}, Phys. Rev. Lett. {\bf 115}, 266601 (2015); S. M. Wu {\it et al.}, {\it ibid.} {\bf 116}, 097204 (2016).
\bibitem{AFM-superfluidity} E. B. Sonin, Solid State Comm. {\bf 25}, 253 (1978); E. B. Sonin, Sov. Phys. JETP {\bf 47}, 1091 (1978).
\bibitem{AFM-T} S. Takei, B. I. Halperin, A. Yacoby and Y. Tserkovnyak, Phys. Rev. B {\bf 90}, 094408 (2014).
\bibitem{STT1} A. S. N\'{u}\~{n}ez, R. A. Duine, P. M. Haney and A. H. MacDonald, Phys. Rev. B {\bf 73}, 214426 (2006); E. V. Gomonay and V. M. Loktev, Low Temp. Phys. {\bf 40}, 17 (2014), and references therein.
\bibitem{STT2} T. Moriyama {\it et al.}, Appl. Phys. Lett. {\bf 106}, 162406 (2015); Y.-W. Oh {\it et al.}, Nat. Nanotechnol. {\bf 11}, 878?884 (2016).
\bibitem{Cheng-PRL2014} R. Cheng, J. Xiao, Q. Niu and A. Brataas, Phys. Rev. Lett. {\bf 113}, 057601 (2014).
\bibitem{Cheng-PRL2016a} R. Cheng, D. Xiao and A. Brataas, Phys. Rev. Lett. {\bf 116}, 207603 (2016).
\bibitem{AFM} A. Auerbach. {\it Interacting Electrons and Quantum Magnetism},  (Springer-Verlag, New York, 1994); S. Sachdev, {\it  Quantum Phase Transitions} (Cambridge University Press, Cambridge, 1999).
\bibitem{Comm1} 
It reads $s=\hbar S L_{t}/\mathcal{V}$, where $L_{t}$ is the thickness of the antiferromagnetic film and $S$, $\mathcal{V}$ are the (dimensionless) spin and volume per site, respectively. 
\bibitem{Fert-PRL1980}
A. Fert and P. M. Levy, Phys. Rev. Lett. {\bf 44}, 1538 (1980).
\bibitem{DMI} I. E. Dzyaloshinskii, Sov. Phys. JETP {\bf 5}, 1259 (1957); T. Moriya, Phys. Rev. {\bf 120}, 91 (1960).
\bibitem{Dzyaloshinskii-JETP1964} I. E. Dzyaloshinskii, Sov. Phys. JETP {\bf 19}, 960 (1964).
\bibitem{Bogdanov-JMMM1994} A. Bogdanov and A. Hubert, J. Magn. Magn. Mater. {\bf 138}, 255 (1994).
\bibitem{yt1}
Note that constructing additional contributions to the free energy in terms of the normalized spin density $\bm{m}$ would only affect the N\'{e}el-order energetics at the subleading order in $J^{-1}(\propto\chi)$, once $\bm{m}$ is integrated out.
\bibitem{SUP} See Supplemental Material for a microscopic calculation of the phenomenological coefficients in the case of a TI substrate, for a brief account of the numerical methods, and for an illustration of two quasi-2D crystal lattices compatible with a homogeneous Dzyaloshinskii-Moriya interaction.
\bibitem{LoM} E. M. Lifshitz and L. P. Pitaevskii, {\it Statistical Physics, Part 2,} 3rd ed., Course of Theoretical Physics, Vol. 9 (Pergamon, Oxford, 1980); E. M. Chudnovsky and J. Tejada. {\it Lectures on Magnetism}, (Rinton Press, New Jersey, 2006). 
\bibitem{Gilbert-1955} T. L. Gilbert, IEEE Trans. Magn. {\bf 40}(6), 3443-3449 (2004).
\bibitem{kimPRB15} S. K. Kim, O. Tchernyshyov, and Y. Tserkovnyak, Phys. Rev. B {\bf 92}, 020402(R) (2015).
\bibitem{Comm4} 
The Gilbert tensor may generally be $\bm{l}$-dependent and anisotropic in spin space. Its dependence on the (normalized) spin density $\bm{m}$ may, however, be disregarded, in the presence of a well-formed N\'{e}el order.
\bibitem{SHE} M. I. Dyakonov and V. I. Perel, JETP Lett. {\bf 13}, 467 (1971); J. E. Hirsch, Phys. Rev. Lett. {\bf 83}, 1834 (1999).
\bibitem{Hals-PRL2011} K. M. D. Hals, Y. Tserkovnyak, and A. Brataas, Phys. Rev. Lett. {\bf 106}, 107206 (2011).
\bibitem{STextT} S. Zhang and Z. Li, Phys. Rev. Lett. {\bf 93}, 127204 (2004); A. Thiaville, Y. Nakatani, J. Miltat, and Y. Suzuki, Europhys. Lett., {\bf 69} (6), 990 (2005); Y. Tserkovnyak, A. Brataas, and G. E. W. Bauer, J. Magn. Magn. Mater. {\bf 320}, 1282 (2008).
\bibitem{Tserkovnyak-PRB2014} Y. Tserkovnyak and S. A. Bender, Phys. Rev. B {\bf 90}, 014428 (2014).
\bibitem{Comm5} Reactive (i.e., nondissipative) torques of the form $\propto(\hat{e}_{z}\times\bm{j})\times\bm{l}$ and $\propto\hspace{-0.05cm}(\bm{j}\cdot\nabla)\bm{l}$ are forbidden due to breaking of the space-group symmetry $\{\bm{l}\rightarrow-\bm{l}\}$. Spin-transfer torques dependent on the spin density $\bm{m}$ are disregarded due to its smallness, $|\bm{m}|\propto J^{-1}$. The (symmetry-allowed) spin-orbit torque $\propto(\bm{l}\cdot\bm{j})(\hat{e}_{z}\times\bm{l})$ is omitted from our treatment but may in principle be present.
\bibitem{Tserkovnyak-RMP2005} Y. Tserkovnyak, A. Brataas and G. E. W. Bauer, Phys. Rev. Lett. {\bf 88}, 117601 (2002); Y. Tserkovnyak, A. Brataas, G. E. W. Bauer, and B. I. Halperin, Rev. Mod. Phys. {\bf 77}, 1375 (2005).
\bibitem{Comm6}
The nonequilibrium description of the heterostructure would generally be complemented with the equation of motion for the charge-current density:\cite{Tserkovnyak-PRB2014} $L\partial_{t}\bm{j}+\hat{\rho}\bm{j}=\bm{E}+\bm{\epsilon}$. Here, $L$ is the self-inductance of the surface, $\hat{\rho}$ is the $2\times2$ resistivity tensor, and $\bm{\epsilon}$ is the motive force induced by the antiferromagnetic dynamics, which is the Onsager-reciprocal of Eq.~\eqref{eq17}. Since our immediate interest in this paper is the dc current-driven dynamics of the insulating antiferromagnet, however, in what follows, we will assume that a certain (fixed) current is injected and maintained within the heavy-metal substrate. 
\bibitem{Hirch} M. W. Hirch, S. Smale and R. L. Devaney. {\it Differential equations, dynamical systems and an introduction to chaos}, (Academic Press, London, 2004). 
\bibitem{Bender-PRL2012} S. A. Bender, R. A. Duine and Y. Tserkovnyak, Phys. Rev. Lett. {\bf 108}, 246601 (2012), see the Supplemental Material.
\bibitem{TI-SOC} Y. Fan {\it et al.}, Nat. Mater. {\bf 13}, 699-704 (2014); A. R. Mellnik {\it et al.}, Nature {\bf 511}, 449-451 (2014).
\bibitem{Cheng-PRL2016b} R. Cheng, J.-G. Zhu and D. Xiao, Phys. Rev. Lett. {\bf 117}, 097202 (2016).
\bibitem{Tserkovnyak-PRB2002} Y. Tserkovnyak, A. Brataas, and G. E. W. Bauer, Phys. Rev. B {\bf 66}, 224403 (2002).
\bibitem{Comm7}
We focus here on the class of insulating antiferromagnets with no intrinsic Dzyaloshinskii-Moriya couplings. Bulk contributions arise, for instance, in $\alpha$-Fe$_{2}$O$_{3}$ (hematite), which exhibits weak ferromagnetism above the Morin temperature $T_{M}=263$ K (with the Dzyaloshinskii vector $\bm{d}$ pointing along the trigonal axis) but no magnetic (texture) superstructure|the space group $D_{3d}^{6}$ is centrosymmetric. On the contrary, FeGe and MnSi have an inhomogeneous Dzyaloshinskii-Moriya coupling (the space group $T^{4}$ is noncentrosymmetric) and can naturally exhibit helicoidal\cite{Dzyaloshinskii-JETP1964} and skyrmion-lattice phases.\cite{Bogdanov-JMMM1994} In these cases, interface-induced Dzyaloshinskii-Moriya interactions may be of a lesser importance.
\bibitem{Fan-NatNano2016} Y. Fan {\it et al.}, Nat. Nanotechnol. {\bf 11} 352-359 (2016).
\bibitem{Comm3}
This description of the surface states requires the Fermi level to lie in the vicinity of the Dirac point, with a possibility of being tuned across it. 
\bibitem{SFN1}
In the same spirit of our phenomenological constructions, we focus on the coupling directly to the N{\'e}el order (whose existence is of course subject to structural symmetries or the relevance of mesoscopic effects). Similar coupling to the net spin density $\bm{m}$, albeit more generic, results in subleading effects for the N{\'e}el order.
\bibitem{ExHam}
I. Garate and M. Franz, Phys. Rev. Lett. {\bf104}, 146802 (2010); K. Nomura and N. Nagaosa, Phys. Rev. B {\bf82}, 161401(R) (2010); Y. Tserkovnyak and D. Loss, Phys. Rev. Lett. {\bf108}, 187201 (2012).
\bibitem{Tserkovnyak-PRBRC2015} Y. Tserkovnyak, D. A. Pesin, and D. Loss, Phys. Rev. B {\bf 91}, 041121(R) (2015).
\end{thebibliography}
\end{document}